\documentclass[]{elsarticle}
\usepackage{euscript,amsmath,amssymb,amsfonts,graphicx,bm}
  \usepackage{paralist}
  \usepackage{epstopdf}
  \usepackage{graphics} 
 \usepackage[colorlinks=true]{hyperref}
 \hypersetup{urlcolor=blue, citecolor=red}
 
\newcommand{\pcb}[1]{\textcolor{black}{#1}}

\numberwithin{equation}{section}

\newcommand{\e}{\mathrm e}

\newcommand{\x}{\mathbf{x}}

\renewcommand{\P}{\mathbb{P}}

\def\E{\mathbb{E}}

\def\P{\mathbb{P}}

\journal{Brain Multiphysics}
\title{Queuing model of axonal transport}
\begin{document}
\begin{frontmatter}

\author{Paul C. Bressloff$^{1}$}
\ead{bressloff@math.utah.edu}

\address{$^1$Department of Mathematics, University of Utah, Salt Lake City, UT 84112 USA}

\begin{abstract}

 The motor-driven intracellular transport of vesicles to synaptic targets in the axons and dendrites of neurons plays a crucial role in normal cell function. Moreover, stimulus-dependent regulation of active transport is an important component of long-term synaptic plasticity, whereas the disruption of vesicular transport can lead to the onset of various neurodegenerative diseases. In this paper we investigate how the discrete and stochastic nature of vesicular transport in axons contributes to fluctuations in the accumulation of resources within synaptic targets. We begin by solving the first passage time problem of a single motor-cargo complex (particle) searching for synaptic targets distributed along a one-dimensional axonal cable. We then use queuing theory
 to analyze the accumulation of synaptic resources under the combined effects of multiple search-and-capture events and degradation. In particular, we determine the steady-state mean and variance of the distribution of synaptic resources along the axon in response to the periodic insertion of particles. The mean distribution recovers the spatially decaying distribution of resources familiar from deterministic population models. However, the discrete nature of vesicular transport can lead to Fano factors that are greater than unity (non-Poissonian) across the array of synapses, resulting in significant fluctuation bursts. We also find that each synaptic Fano factor is independent of the rate of particle insertion but increases monotonically with the amount of protein cargo in each vesicle. This implies that fluctuations can be reduced by increasing the injection rate while decreasing the cargo load of each vesicle.

 \end{abstract}


\end{frontmatter}

\section{Introduction}
\label{sec7:axon}

Axons of neurons can extend up to 1m in large organisms but synthesis of many of their components occurs in the cell body. The healthy growth and maintenance of an axon depends on the interplay between the axonal cytoskeleton and the active transport of various organelles
and macromolecular proteins along the cytoskeleton \cite{Hirokawa05,Brown13,Maday14,Maeder14a}. The disruption of axonal transport occurs in many neurodegenerative diseases, including
Alzheimer's disease, Parkinson's disease, amyotrophic lateral sclerosis (also known
as Lou Gherig's disease), and Huntington's disease \cite{Vos08,Millecamps13}. All of these diseases exhibit an aberrant accumulation
of certain cellular components and excessive focal swelling of the axon, ultimately leading to axon degeneration.

\begin{figure}[t!]
  \centering
  \includegraphics[width=10cm]{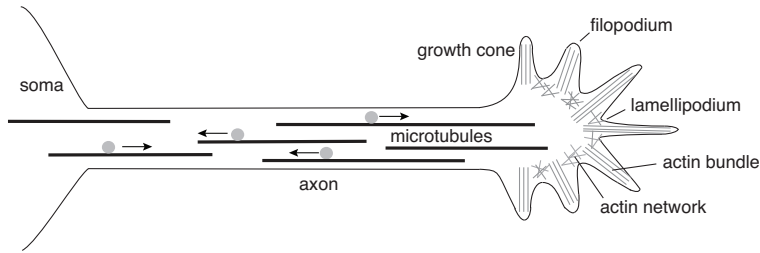}
  \caption{Bidirectional transport of intracellular cargo along an overlapping 1D array of microtubules
within an axon.}
  \label{1Daxon}
\end{figure}

The axonal cytoskeleton contains microtubules and actin microfilaments, which play a role in long-range and short-range axonal transport, respectively, and neurofilaments that provide structural support for the axon.
Actin microfilaments are mainly found
beneath the axon membrane, forming evenly-spaced ring-like structures that wrap
around the circumference of the axon shaft. They are also enriched in growth cones and axon terminals. Actin microfilaments tend to be involved in more short-range transport, such as the transfer of organelles and proteins from microtubules to targets in the membrane via myosin molecular motors. Longer-range vesicular transport involves microtubules, which are polarized polymers with biophysically distinct ($+$) and $(-)$ ends. This polarity determines the preferred direction in which an individual molecular motor moves. For example, kinesin moves towards the $(+)$ end whereas dynein moves towards the $(-)$ end of a microtubule. It turns out that microtubules align axially along an axon, with plus
ends pointing away from the cell body. They do not extend over the whole length
of an axon, having typical lengths of around $100\, \mu$m, but rather form an overlapping array from the cell body to the axon terminal, see Fig. \ref{1Daxon}. Individual vesicles are often transported by multiple motors forming a motor/cargo complex. The velocity state of the complex then depends on the current number of kinesin and/or dynein motors bound to a microtubule. The resulting tug-of-war between opposing motors can result in random intermittent behavior, with constant velocity movement in both directions along the microtubular array (bidirectional transport), interrupted by brief pauses or fast oscillatory movements that may correspond to 
localization at specific targets such as synapses or the growth cone at the axon terminal \cite{Gross04,Welte04,Klumpp05,Muller08,Muller08a,Newby10a,Newby10c}. Analogous behavior has been observed during the transport of mRNA in dendrites and oocytes \cite{Rook00,Dynes07,Ciocanel17}. There are also higher-dimensional versions of motor-driven transport within the
soma of neurons and in most non--polarized animal cells, which involves the microtubular network that projects radially from organizing centers known as centrosomes \cite{Burute19}. 

Axonal transport is typically divided into two main categories based upon the observed speed \cite{Brown03,Brown13}: fast transport ($1-9 \, \mu$m/s) of organelles and vesicles and slow transport ($0.004-0.6 \, \mu$m/s) of soluble proteins and cytoskeletal elements.  Slow transport is further divided into two groups; actin and actin-bound proteins are transported in slow component A while cytoskeletal 
polymers such as microtubules and neurofilaments are transported in slow component B.
It had originally been assumed that the differences between fast and slow components were due to differences in transport mechanisms, but direct experimental observations now indicate that they all involve fast 
motors but differ in how the motors are regulated. Membranous organelles such as mitochondria and vesicles, which function primarily to deliver membrane and protein components to sites along the axon and at the axon tip, move rapidly in a 
unidirectional or bidirectional manner, pausing only briefly. In other words, they have a high duty ratio -- the proportion of time a cargo complex is actually moving. On the other hand, cytoskeletal polymers such as neurofilaments
move in an intermittent and bidirectional manner, pausing more often and for longer time intervals; such transport has a low duty ratio.

When modeling the active transport of intracellular cargo over relatively long distances, it is often convenient to ignore the microscopic details of how individual motors perform a single step (as described by  
Brownian ratchet models for example \cite{Reimann02}), and to focus instead on the transitions between the different velocity states as described by a velocity jump process \cite{Bressloff13,Xue17}. The corresponding differential Chapman-Kolmogorov (CK) equation for the probability density is often approximated by a Fokker-Planck equation using a quasi-steady-state reduction \cite{reed1990,Smith01,Friedman05,Newby10a,Ciocanel17}. (Alternatively, the motion of each motor can be modeled directly in terms of a stochastic differential equation \cite{McKinley12}.) Velocity jump processes have also been used to model slow axonal transport, in which the slow rate of movement of a population is an average of rapid bidirectional movements interrupted 
by prolonged pauses (stop-and-go hypothesis) \cite{Jung09,Li12}.
Given a stochastic model for the motion of an individual motor/cargo complex (particle), one can formulate the transport and delivery of a vesicle to some cellular target as a classical search-and-capture process. That is, given the initial position of the particle, one can determine the first passage time (FPT) distribution for the particle to be absorbed by the target and calculate various moments such as the mean FPT (MFPT). One issue of interest is how to optimize the search process (minimize the MFPT) with respect to the transition rates between the different velocity states, 
which is a major feature of so-called random intermittent search processes \cite{Loverdo08,Benichou10,Benichou11,Bressloff11,Bressloff13}. In the case of multiple independent searchers one can also consider the FPT of the fastest particle to find a target, which is an example of an extreme statistic \cite{Fortin15,Schuss19,Lawley20}.

\pcb{In the case of multiple, non-interacting motor particles one can model axonal transport in terms of an advection-diffusion equation for the concentration of particles along the axon, which is the analog of the Fokker-Planck equation at the single-particle level. This type of population model has been used extensively to study the problem of axonal transport within the context of axonal growth \cite{Graham06,McLean06,OToole08,Zadeh10,OToole11,Diehl14}. Such studies typically focus on the transport and delivery of tubulin (the basic monomeric unit of microtubules) to the growth cone at the axon terminal. This determines the rate of microtubule polymerization within the growth cone and thus the speed of axonal elongation. (A complicating factor from a mathematical perspective is that one has to deal with a moving boundary value problem.) Population models provide a good framework for studying axonal growth because there is a continuous flux of tubulin at the axon terminal such that stochastic effects can be ignored. However, the discrete and stochastic nature of vesicular transport and delivery to individual synaptic targets is much more significant, and cannot be accounted for using population models. Sources of noise include the random motion of individual motor complexes along the axon, the stochastic nature of particle injection and capture, and resource degradation. This target-centric perspective motivates the construction and analysis of discrete particle models, which is the focus of this paper. 
}

\pcb{Our main goal is to analyze the stochastic accumulation of resources in one or more synaptic targets due to the active transport and delivery of vesicles by multiple motor/cargo complexes (multiparticle search-and-capture).} Each time a complex is captured by a synaptic target, it secretes a vesicle containing a fixed amount of resources (eg. proteins), which we refer to as a burst event. Following target capture, the complex is either sorted for degradation or recycled for another round of transport and delivery. The random sequence of burst events under multiple rounds of search-and-capture leads to an accumulation of resources within a target, which is counteracted by subsequent degradation. (For simplicity, we lump together all downstream processes that `use up' the supplied resources.)  At the multiple particle level, the accumulation of resources will also depend on the rule for injecting new particles into the axon from the soma. We will assume that particles are inserted sequentially, either at periodic intervals $\Delta_0$ or according to a renewal process with waiting time density $\psi(\tau)$. An alternative injection protocol would be to assume multiple particles are simultaneously injected into the axon and after each particle has delivered its cargo, it returns to the soma where it is resupplied with resources after some delay. However, this is based on the unrealistic assumption that there is a fixed number of particles. 

As we have recently highlighted elsewhere \cite{Bressloff20}, there are interesting parallels between axonal transport and cytoneme-based morphogenesis in invertebrates. More specifically, there is growing experimental evidence for a direct cell-to-cell signaling mechanism during embryonic development, which involves the active transport of morphogenic receptors or ligands along cytonemes, which are thin, actin-rich cellular extensions with a diameter of around 100 nm and lengths that vary from 1 to 200 $\mu$m \cite{Ramirez99,Roy11,Gradilla13,Kornberg14}. Each cytoneme can be treated as a tunneling nanotube linking a source cell and a target cell, along which vesicles are actively transported by myosin motors. Since the steady-state amount of resources in a target cell is an exponentially decreasing function of cytoneme length, this provides a mechanism for the formation of a morphogen gradient \cite{Teimouri16,Bressloff18,Kim19}.\footnote{Cytoneme-based morphogenesis in vertebrates such as zebrafish appears to involve a different transport mechanism \cite{Sanders13,Kornberg14,Stanganello16}. In these systems, morphogen is located at the tip of a cytoneme growing out from a source cell. When the tip makes contact with a target cell, it delivers its cargo and then rapidly retracts back to the source cell. The cytoneme then renucleates from the source cell and initiates a new round of search-and-capture. Such a process can be modeled in terms of a single particle (cytoneme tip) executing multiple rounds of search-and-capture \cite{Bressloff19,Bressloff20}.}
Analogous to the accumulation of morphogen in a target cell due to active transport along a cytoneme, we show how the accumulation of synaptic resources in response to axonal transport can be modeled as an infinite server queue \cite{Takacs62,Liu90}. 

Queuing theory concerns the mathematical analysis of waiting lines formed by customers randomly arriving at some service station and staying in the system until they receive service from a group of servers. The multiparticle search-and-capture model is mapped into a queuing process as follows: The delivery of a vesicle of size $C$ represents the arrival of $C$ customers, a given target represents the service station, and the degradation of resources is the analog of customers exiting the system after service. Since the resource elements are degraded independently of each other, the effective number of servers in the corresponding queuing model is infinite. The distribution $F(t)$ of customer interarrival times is determined by the first passage time distributions of the individual particles and the times at which they initiate their searches. (In the case of axonal transport, the latter would depend on the rate at which motor complexes enter the axon from the soma.) Similarly, the service time distribution $H(t)$ is determined by the degradation of resources, which is taken to be a Poisson process. It follows that the model maps to a $G/M/\infty$ queue. Here the symbol $G$ denotes a general interarrival time distribution, the symbol $M$ stands for a Markovian service time distribution, and $\infty$ denotes an infinite number of servers. The advantage of mapping the stochastic process to a $G/M/\infty$ queue is that one can use renewal theory to determine the moments of the steady-state number of resources within a target.

The structure of the paper is as follows. In section 2 we introduce the basic axonal transport model. We begin in section 2.1 by briefly considering a population version of the model that determines the evolution of the concentration of motor particles along a one-dimensional axon, under the combined effects of advection-diffusion and absorption by synaptic targets. \pcb{As we previously noted, although such a model captures the macroscopic distribution of resources along an axon, it cannot account for the discrete and stochastic nature of resource accumulation within an individual synapse. Therefore, in section 2.2} we turn to a stochastic model of a single particle and solve the resulting inhomogeneous Fokker-Planck equation using Laplace transforms and Green's functions. The solution for the probability flux into a given synaptic target is then used to derive expressions for the splitting probability and conditional mean first passage time (MFPT) for the target to capture the particle. In section 3 we extend the single-particle analysis to the case of multiple particles injected sequentially into the axon in order to determine the accumulation of synaptic resources due to competition between the transport/delivery of cargo and degradation. In particular, we show how the statistics of a single search-and-capture model can be incorporated into an infinite server queue, where motor particles represent customer batches, synapses correspond to service stations, and degradation signals the exit of a customer. We then use queuing theory to construct a renewal equation for the Binomial moments of the number of
resources in each target. In section 4 we use the renewal equation to derive expressions for the steady-state mean and variance of the distribution of synaptic resources, and explore the parameter dependence of the fluctuations. Possible extensions of the analysis are described in section 5.

\section{Axonal transport model}

\begin{figure}[t!]
\raggedleft
\includegraphics[width=8cm]{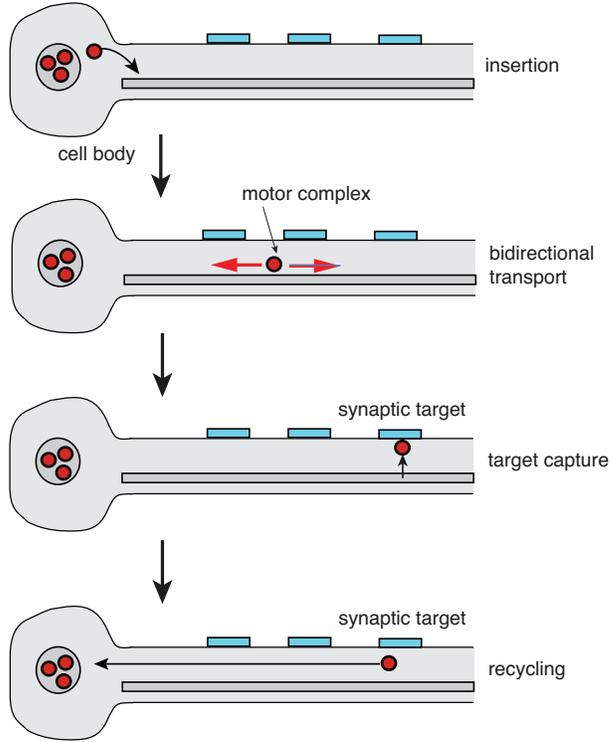} 
\caption{Model of axonal transport. \pcb{The axon is treated as a cable of length $L_T$ with partially absorbing synaptic targets distributed along a subregion of length $L$, $L<L_T$.} (i) Particles from a compartmental pool are inserted into the axon at a mean rate $J_0$. (ii) Each particle undergoes bidirectional transport along the axon until it is captured by the $k$-th target with splitting probability $\pi_k$, and secretes a discrete packet of resources (burst event). (iii) Following target capture, the particle is either sorted for degradation or recycled to the particle pool.}
\label{fig2}
\end{figure}

A schematic illustration of our basic model of axonal transport is shown in Fig. \ref{fig2}. \pcb{For simplicity, we treat the axon as a finite cable of length $L_T$ with a pool of motor-cargo complexes (particles) located at the end $x=0$ and a set of {\em en passant} synapses located in the subregion $x\in [0,L]$ with $L < L_T$.} Particles are inserted into the axon at a mean rate $J_0$. Each time a particle enters the axon, it executes a stochastic search for a synaptic target. When a particle is within a neighborhood of a target, it can be captured at some rate $\kappa$. Following target capture, the particle secretes a discrete packet (vesicle) of $C$ resources (eg. proteins), which we refer to as a burst event, after which it is either sorted for degradation or recycled to the particle pool. The random sequence of burst events under multiple rounds of search-and-capture leads to an accumulation of resources within the synaptic target, which is counteracted by degradation at some rate $\gamma$. The main elements of the model are the dynamics of the motor-cargo complexes along the axon, and the rules of particle insertion, capture and recycling.

\subsection{Population model}

The simplest version of the axon transport model is to consider a population of motor particles within the axon and to assume a uniform distribution of synaptic targets with density $\rho_0$. Let $c(x,t)$ be the concentration of particles at position $x$ along the axon at time $t$, which evolves according to the advection-diffusion equation
\begin{equation}
 \label{adv}
\frac{\partial c}{\partial t}=-v\frac{\partial c}{\partial x}+D\frac{\partial^2 c}{\partial x^2}-\kappa \rho_0 c,\quad t >0,\quad 0<x<L,
\end{equation}
where $\kappa$ is the target absorption rate (in units of velocity), $v$ is an effective drift velocity and $D$ is an effective diffusivity. Equation (\ref{master}) is supplemented by the boundary conditions $J(0,t)=J_0$ and $c(L,t)=0$, where
\begin{equation}
J(x,t)=vc(x,t)-D\frac{\partial c}{\partial x}(x,t).
\end{equation}
\pcb{The absorbing boundary condition at $x=L$ implies that if a motor particle travels beyond the region containing the {\em en passant} synapses, then it either delivers its cargo to some other target (such as the growth cone at $x=L_T$) or simply degrades and returns to the motor pool. One could consider a more general Robin boundary condition at $x=L$, in which there is a non-zero probability that the particle is reflected and then subsequently absorbed by one of the synaptic targets. The precise choice of boundary condition does not affect the main results developed in this paper.}

\begin{figure}[b!]
\raggedleft
\includegraphics[width=8cm]{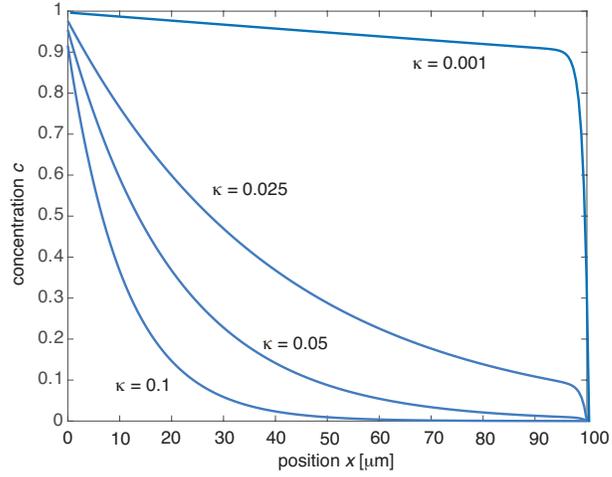} 
\caption{Population model. Plot of steady-state concentration $c(x)$ along the axon for different absorption rates $\kappa$. Other parameter values are $D=v=\rho_0=1$ and $L=100$. The injection rate is $J_0=1$.}
\label{fig3}
\end{figure}

The absorption of motor particles by the synaptic targets leads to a build up of synaptic resources that is counteracted by degradation. Taking $n(x,t)$, $x\in [0,L]$, to be the concentration of synaptic resources along the axon then
\begin{equation}
\label{CD2}
\frac{\partial n}{\partial t} = \pcb{\kappa \rho_0 c} -\gamma n,
\end{equation}
where $\gamma$ is the degradation rate. \pcb{(Note that there is no diffusion term in the above equation because the synaptic resources are localized within discrete compartments.)} 
The steady-state solution is of the form
\begin{equation}
c(x)=A_+\e^{\mu_+ x}+A_-\e^{\mu_- x},\quad n(x)=\pcb{\frac{\kappa \rho_0c(x)}{\gamma}},
\end{equation}
with
\begin{equation}
\mu_{\pm}=\frac{1}{2D}\left [v\pm \sqrt{v^2+4\kappa \rho_0 D}\right ].
\end{equation}
Imposing the boundary conditions generates the constraints
\begin{align}
\mu_- A_++\mu_+ A_-=J_0,\quad A_+\e^{\mu_+ L}+ A_-\e^{\mu_-L}=0.
\end{align}
Hence,
\begin{align}
c(x)=\frac{J_0}{\mu_-\e^{-\mu_+L}-\mu_+\e^{-\mu_-L}}\left \{ \e^{-\mu_+(L-x)}-\e^{-\mu_-(L-x)} \right \}.
\end{align}
In Fig. \ref{fig3} we plot the steady-state concentration $c(x)$ as a function of $x$ for various absorption rates $\kappa$. Units of length and time are $\mu$m and seconds, respectively. It can be seen that if $\kappa /v$ is not too small then the concentration profile in the bulk of the domain is an exponentially decreasing function of $x$. However, as $\kappa/v$ decreases, a boundary layer develops at the distal end $x=L$. \pcb{(A boundary layer at $x=L$ would also occur in the case of a reflecting boundary, for example, but now there would be an increase in the concentration within the boundary layer.)} For sufficiently small $\kappa/v$, the bulk concentration is approximately uniform but comes at the cost of a slow build up of resources within the synapses.

\pcb{One mechanism for generating a more uniform distribution of resources for relatively fast absorption is to allow for the reversible delivery of vesicles, which has been observed experimentally in {\em{C. elegans}} and {\em{Drosophila}} \cite{Wong12,Maeder14,Maeder14a} and demonstrated theoretically using a generalized version of the population model (\ref{adv}) that keeps track of motor-complexes that are no longer carrying a vesicle \cite{Bressloff15,Karamched17}. More specifically, let $c_1(x,t)$  and $c_0(x,t)$ denote the density of motor-complexes with and without an attached vesicle, respectively, and denote the forward and backward rates for cargo delivery by $\kappa_{\pm}$. The transport of each motor population is described by an advection-diffusion equation, but we now include the transitions between the two populations due to the reversible exchange of vesicles with synaptic targets. Thus, equation (\ref{adv}) becomes
\begin{subequations}
\label{master}
\begin{align}
\frac{\partial c_0}{\partial t}&=-v_0\frac{\partial c_0}{\partial x}+D\frac{\partial^2c_0}{\partial x^2}
-\kappa_-n c_{0}+\kappa_+\rho_0 c_1,\\
\frac{\partial c_1}{\partial t}&=-v_1\frac{\partial c_1}{\partial x}+D\frac{\partial^2c_1}{\partial x^2}+\kappa_-n\, u_{0}-\kappa_+ \rho_0u_1
\end{align}
\end{subequations}
with $c_k(L,t)=0$, $k=0,1$. We are allowing for the possibility that the mean speed of a motor complex may differ, depending on whether or not it is bound to a vesicle. It is also assumed that motor-complexes with and without cargo are injected at the somatic end $x=0$ at constant rates $J_1$, and $J_0$, respectively. 
Finally, in order to incorporate the reversible exchange between motor complexes and synaptic targets, it is necessary to modify equation (\ref{CD2}) according to 
\begin{align}
\frac{\partial n}{\partial t}&= \kappa_+ \rho_0 c_1(x,t)-\kappa_-n(x,t)c_0(x,t)-\gamma n(x,t).
\label{c}
\end{align}
For the sake of illustration, suppose that $\gamma=0$ and $L\rightarrow \infty$. The steady-state resource distribution is then
\begin{equation}
\label{c2}
n(x)=\frac{\kappa_+c_1(x)}{\kappa_-c_0(x)},
\end{equation}
with
\begin{align}
u_j(x)=\frac{J_j\e^{-x/\xi_j}}{ D/\xi_j+v_j},\quad \xi_j=\frac{2D}{-v_j+ \sqrt{v_j^2+4D \gamma_u}}
\end{align}
for $j=0,1$. Combining with equation (\ref{c2}) then yields the following result for the steady-state density of synaptic vesicles:
\begin{equation}
n(x)=\frac{\kappa_+}{\kappa_-}\frac{J_1}{J_0}\frac{ D/\xi_0+v_0}{ D/\xi_1+v_1}\e^{-\Gamma x},
\end{equation}
where $\Gamma= \xi_1^{-1}-\xi_0^{-1}.$ In particular, if the transport properties of the motor-complex are independent of whether or not it is bound to a vesicle ($v_0=v_1$), then $\xi_0=\xi_1$ and we have a uniform vesicle distribution $n(x)=(\kappa_+/\kappa_-)(J_1/J_0)$.  
}

\subsection{Single particle model}

\pcb{As we highlighted in the introduction, the discrete and stochastic nature of vesicular transport and delivery to individual synaptic targets cannot be accounted for using population models. This motivates the construction and analysis of discrete particle models. Here we will consider stochasticity at the single particle level. We will then show in sections 3 and 4 how the statistics of single-particle search can be incorporated into a multiple particle model using queuing theory, assuming that injected particles are identical and noninteracting so that each particle independently searches for a target according to the same stochastic process.}

Suppose that a particle is injected into the axon at time $t=0$. Let $p(x,t)$ denote the probability density that the given particle is at position $x$ at time $t$, having started at $x(0)=0$ with $t\geq 0$. In the absence of any synaptic targets, $p$ evolves according to the Fokker-Planck equation
\begin{equation}
 \label{master}
\frac{\partial p}{\partial t}=-v\frac{\partial p}{\partial x}+D\frac{\partial^2 p}{\partial x^2}\equiv -\frac{\partial J}{\partial x},\quad t >0,\quad 0<x<L,
\end{equation}
where $J$ is the probability flux. Equation (\ref{master}) is supplemented by the boundary conditions $J(0,t)=0=p(L,t)$ and the initial condition $p(x,0)=\delta(x)$. \pcb{Again the absorbing boundary condition at $x=L$ is based on the assumption that once the motor particle crosses the right-hand boundary it delivers its cargo to some other region of the axon beyond $x=L$ or degrades.} Note that equation (\ref{master}) can be derived from a more detailed biophysical model that takes the form of a velocity jump process. The latter assumes that the particle randomly switches between different velocity states according to some irreducible Markov chain. In the limit of fast switching one can obtain (\ref{master}) using an adiabatic approximation \cite{Newby10a}. 

Now suppose that rather than a uniform distribution of synapses, there exists a finite set of synaptic targets located at positions $x_k\in (0,L)$ along the axon, $k=1,\ldots,M$. In the one-dimensional case, the synapses can be represented as point-like sinks so that equation (\ref{master}) becomes
\begin{equation}
 \label{master2}
\frac{\partial p}{\partial t}=-v\frac{\partial p}{\partial x}+D\frac{\partial^2 p}{\partial x^2}-\kappa \sum_{k=1}^M\delta(x-x_k)p,\quad t >0,\quad 0<x <L,
\end{equation}
where $\kappa$ is again the rate of absorption (in units of velocity). 
Next we introduce the survival probability that the particle hasn't been absorbed by a target in the time interval $[0,t]$:
\begin{equation}
\label{Q0}
Q(t)=\int_{0}^{L}p(x,t)dx.
\end{equation}
Differentiating both sides with respect to $t$ and using equation (\ref{master2}) implies that
\begin{align}
\label{QJ}
\frac{dQ(t)}{dt}&=-\int_{0}^{L}\left [\frac{\partial J}{\partial x}+\kappa \sum_{k=1}^M\delta(x-x_k)p \right ]dx=-\sum_{k=1}^MJ_k(t)-J_L(t),
\end{align}
where $J_L(t)=J(L,t)$ is the probability flux at the distal end of the axon and
\begin{equation}
J_k(t)=\kappa  p(x_k,t)
\end{equation}
is the probability flux into the $k$-th target. Let ${\mathcal T}_k$ denote the FPT that the particle is captured by the $k$-th target, with ${\mathcal T}_k=\infty$ indicating that it is not captured. The splitting probability that the particle is captured by the $k$-th target is
\begin{align}
\label{pi}
\pi_k &:= \mathbb{P}[0< {\mathcal T}_k<\infty] = \int_0^\infty J_k(t') dt' =\widetilde{J}_k(0)=\kappa\widetilde{p}(x_k,0),
\end{align}
where $\widetilde{J}_k(s)$ is the Laplace transform of $J_k(t)$ etc.
We have used the fact that for the given class of partially absorbing targets, $J_k(t)$ is equivalent to the conditional FPT density.
The corresponding conditional MFPT is defined by
\begin{equation}
T_k=\mathbb{E}[{\mathcal T}_k | {\mathcal T}_k < \infty].
\end{equation}
It follows that
the conditional MFPT is given by 
\begin{align}
\pi_k T_k &= \int_0^\infty t J_k(t)dt =-\widetilde{J}_k'(0)=-\kappa \widetilde{p}'(x_k,0).
\label{piT}
\end{align}
Similarly, the second order moments of the FPT density are
\begin{align}
\pi_k T^{(2)}_k &= \int_0^\infty t^2 J_k(t)dt =\widetilde{J}_k''(0)=\kappa \widetilde{p}''(x_k,0).
\label{piT2}
\end{align}

\pcb{Integrating both sides of equation (\ref{QJ}) with respect to $t$ after imposing the conditions $Q(0)=1$ and $Q(t)\rightarrow 0$ as $t\rightarrow \infty$, we have
\begin{equation}
1=\sum_{k=1}^M \int_0^{\infty} J_k(t)dt+\int_0^{\infty}J_L(t)dt = \sum_{k=1}^M \widetilde{J}_k(0)+\widetilde{J}_L(0).
\end{equation}
Equation (\ref{pi}) then implies that
\begin{equation}
\sum_{k=1}^M \pi_k=1-\widetilde{J}_L(0)<1.
\end{equation}
The physical interpretation of this result is that the total probability that the particle is absorbed by one of the $M$ synaptic targets is less than unity due to the fact that there is a nonzero probability of absorption at the right-hand boundary $x=L$. If the right-hand boundary had been reflecting, then $\widetilde{J}_L(0)\equiv 0$ and $\sum_{k=1}^M\pi_k =1$.
}

Equations (\ref{pi}) and (\ref{piT}) imply that the splitting probabilities and conditional MFPTs can be determined by solving equation (\ref{master2}) in Laplace space:
\begin{equation}
 \label{masterLT}
s\widetilde{p}(x,s)-\delta(x) =-v\frac{\partial \widetilde{p}}{\partial x}+D\frac{\partial^2 \widetilde{p}}{\partial x^2}-\kappa \sum_{k=1}^M\delta(x-x_k)\widetilde{p},\quad 0<x<L.
\end{equation}
This is supplemented by the boundary conditions
\[\widetilde{J}(0,s)=0,\quad \widetilde{p}(L,s)=0.\]
Introducing the Green's function $G(x,s|x_0)$ according to
\begin{subequations}
 \label{G}
 \begin{equation}
-v\frac{\partial {G}(x,s|x_0)}{\partial x}+D\frac{\partial^2 {G}(x,s|x_0)}{\partial x^2}- s {G}(x,s|x_0)= -\delta(x-x_0),\quad 0<x<L,
\end{equation}
with 
\begin{equation}
vG(0,s|x_0)-D\left . \frac{\partial {G}(x,s|x_0)}{\partial x}\right |_{x=0}=0 ,\ G(L,s|x_0)=0,
\end{equation}
\end{subequations}
we can formally write the solution as
\begin{equation}
\widetilde{p}(x,s)=G(x,s|0)-\kappa \sum_{l=1}^M G(x,s|x_l)\widetilde{p}_l(s),
\end{equation}
where $\widetilde{p}_k(s)=\widetilde{p}(x_k,s)$. Finally, the unknown functions $\widetilde{p}_k(s)$ are obtained by setting $x=x_k$ to yield the following matrix equation for the target fluxes
\begin{equation}
\label{matrix}
\sum_{l=1}^{M}(\delta _{k,l}+\kappa G(x_k,s|x_l))\widetilde{J}_l(s)=\kappa G(x_k,s|0).
\end{equation}
   
 One can derive an explicit expression for the Green's functions $G$, which is given by
 \begin{subequations}
 \label{Gv}
\begin{equation}
G(x,s|x_0)=A^{-1}\left (\frac{\e^{\lambda_-(s)x}}{\lambda_+(s)}-\frac{\e^{\lambda_+(s)x}}{\lambda_-(s)}\right )\left (\e^{\lambda_-(s)(x_0-L)}-\e^{\lambda_+(s)(x_0-L)}\right),\ x< x_0,
\end{equation}
and
\begin{equation}
G(x,s|x_0)=A^{-1}\left (\frac{\e^{\lambda_-(s)x_0}}{\lambda_+(s)}-\frac{\e^{\lambda_+(s)x_0}}{\lambda_-(s)}\right )\left (\e^{\lambda_-(s)(x-L)}-\e^{\lambda_+(s)(x-L)}\right),\  x>x_0,
\end{equation}
\end{subequations}
where
\begin{equation}
A= D\left [\left (1-\frac{\lambda_+}{\lambda_-}\right )\e^{-\lambda_-L}+\left (1-\frac{\lambda_-}{\lambda_+}\right )\e^{-\lambda_+L}\right ]\e^{(\lambda_++\lambda_-)x_0}
\end{equation}
and
\begin{equation}
\lambda_{\pm}(s)=\frac{1}{2D}\left [v\pm \sqrt{v^2+4s D}\right ].
\end{equation}
$G$ takes a particularly simple form in the case of pure diffusion ($v=0$), namely, it is given by the Green's function of the modified Helmholtz equation:
\begin{subequations}
\label{GD}
\begin{equation}
G(x,s|x_0)=\frac{1}{\sqrt{sD}}\frac{\cosh(\sqrt{s/D}x)\sinh(\sqrt{s/D}(L-x_0))}{\cosh(\sqrt{s/D}L)},\quad x< x_0,
\end{equation}
and
\begin{equation}
G(x,s|x_0)=\frac{1}{\sqrt{sD}}\frac{\cosh(\sqrt{s/D}x_0)\sinh(\sqrt{s/D}(L-x))}{\cosh(\sqrt{s/D}L)},\quad x>x_0.
\end{equation}
\end{subequations}

 In order to determine the splitting probabilities $\pi_k$ and MFPTs $T_k$ it is necessary to determine $\widetilde{J}_k(s)$ in the small $s$ limit. This requires taking into account the fact that the Green's function can be expanded as
\begin{equation}
\label{sexp}
G(x,s|x_0)=G_0(x|x_0)+sG_1(x|x_0)+O(s^2),
\end{equation}
where 
\begin{equation}
 \label{G0}
G_0(x|x_0)=\lim_{s\rightarrow 0}G(x,s|x_0),\quad G_1(x|x_0)=\lim_{s\rightarrow 0}\frac{d}{ds}G(x,s|x_0). \end{equation}
Substituting the series expansion (\ref{sexp}) into the matrix equation (\ref{matrix}) and expanding the fluxes according to
\begin{equation}
\widetilde{J}_l(s)=\pi_l(1-sT_l+O(s^2))
\end{equation}
gives
\begin{align}
\label{matrix2}
&\sum_{l=1}^{M}\left (\delta _{k,l}+\kappa G_{0}(x_k|x_l)+s\kappa  G_{1}(x_k|x_l)+O(s^2)\right ) \\
&\times \pi_l(1-sT_l+s^2T_l^{(2)}/2+\ldots) =\kappa \left (G_{0}(x_k|0)+s G_{1}(x_k|0)+O(s^2)\right ). \nonumber
\end{align}
Collecting $O(1)$ terms yields the equation
\begin{align}
\label{hier0}
\sum_{l=1}^{M}\left (\delta _{k,l}+\kappa G_{0}(x_k|x_l)\right )\pi_l=\kappa G_0(x_k|0).
\end{align}
Introducing the matrix ${\bf A}(\kappa)$ with elements
\begin{equation}
A_{kl}(\kappa)=\delta _{k,l}+\kappa G_{0}(x_k|x_l),
\end{equation}
which is invertible for $\kappa/v <1$, we have the solution
\begin{equation}
\label{pik}
\pi_k=\kappa \sum_{l=1}^MA_{kl}^{-1}(\kappa)G_0(x_l|0).
\end{equation}
The conditional MFPTs are obtained by collecting $O(s)$ terms in equation (\ref{matrix2}):
\begin{align}
\label{hier1}
\sum_{l=1}^{M}\left (\delta _{k,l}+\kappa G_{0}(x_k|x_l)\right )\pi_lT_l&=\kappa \sum_{l=1}^{M}G_1(x_k|x_l)\pi_l-\kappa G_1(x_k|0).
\end{align}
It follows that
\begin{equation}
\label{Tk}
\pi_kT_k=\kappa \sum_{l=1}^MA_{kl}^{-1}(\kappa) \left (  \sum_{l'=1}^{M}G_1(x_l|x_{l'})\pi_{l'}- G_1(x_l|0)\right ).
\end{equation}

\begin{figure}[t!]
\raggedleft
\includegraphics[width=8cm]{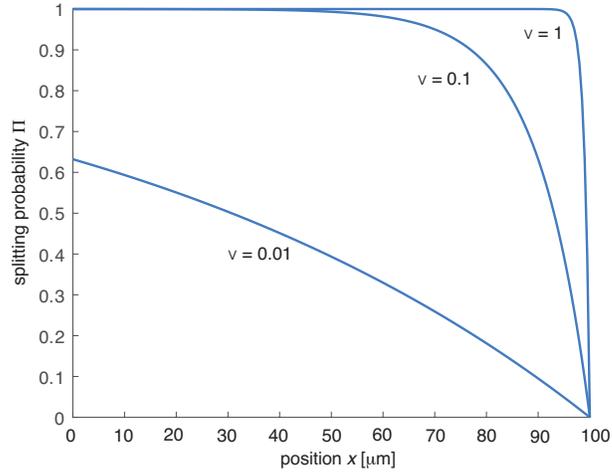} 
\caption{Single particle model. Plot of splitting probability $\Pi(x)$ along the axon for different drift velocities $v$ and fixed $\epsilon =\kappa/v\ll 1$. Other parameter values are $D=1$ and $L=100$.}
\label{fig4}
\end{figure}

\begin{figure}[b!]
\raggedleft
\includegraphics[width=8cm]{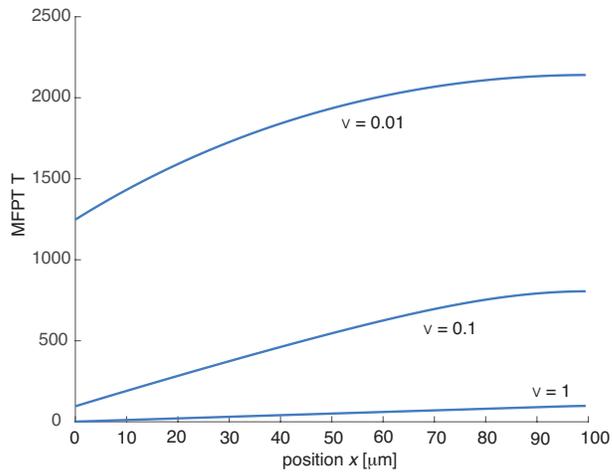} 
\caption{Single particle model. Plot of MFPT $T(x)$ along the axon for different drift velocities $v$ and fixed $\epsilon =\kappa/v\ll 1$. Other parameter values are $D=1$ and $L=100$.}
\label{fig5}
\end{figure}

Since the Green's function is $O(1/v)$, it follows that for $\epsilon \equiv\kappa/v\ll 1$ (slow absorption) we can carry out a perturbation expansion in $\epsilon$, which yields the leading order expressions 
\begin{equation}
\pi_k=\epsilon vG_0(x_k|0)+O(\epsilon^2),
\end{equation}
and
\begin{equation}
T_k=-\frac{G_1(x_k|0)}{G_0(x_k|0)}+O(\epsilon).
\end{equation}
Note that $G_1(x_k|0)<0$. Hence, in the slow absorption regime, the splitting probabilities and conditional MFPTs of each synaptic target are approximately independent of the locations of the other synapses. We can then define the functions
\begin{equation}
\Pi(x)= vG_0(x|0),\quad T(x)=-\frac{G_1(x|0)}{G_0(x|0)},
\end{equation}
such that $\pi_k\approx \Pi(x_k)/\epsilon$ and $T_k\approx T(x_k)$. Example plots of the functions $\Pi(x)$ and $T(x)$ are shown in Figs. \ref{fig4} and \ref{fig5}, respectively. In particular, the splitting probability is monotonically decreasing function of position along the axon, consistent with the population model for small $\kappa/v$. As expected, the MFPT increases distally along the axon and is a decreasing function of the drift velocity $v$.

Outside the slow absorption regime, we can determine $\pi_k$ and $T_k$ by inverting the matrix ${\bf A}(\kappa)$. For illustrative purposes, we focus on the case of two synaptic targets at positions $x_1$ and $x_2$, respectively. Inverting the matrix
\begin{equation}
{\bf A}=\left (\begin{array}{cc} 1+\kappa G_{11}& \kappa G_{12} \\
\kappa G_{21} & 1+\kappa G_{22}\end{array} \right ),
\end{equation}
with $G_{kl}=G_0(x_k|x_l)$, $G_{k}=G_0(x_k|0)$ and setting $N=2$ in equation (\ref{pik}) yields
\begin{subequations}
\begin{align}
\pi_1&=\kappa {\mathcal D}^{-1} \left ([1+\kappa G_{22}]G_1 -\kappa G_{12} G_2\right ),\\
\pi_2&=\kappa {\mathcal D}^{-1} \left ([1+\kappa G_{11}]G_2 -\kappa G_{21} G_1\right ).
\end{align}
\end{subequations}
Here
\begin{equation}
{\mathcal D}=\mbox{det}{\bf A}(\kappa)=1+\kappa(G_{11}+G_{22})+\kappa^2(G_{11}G_{22}-G_{12}G_{21}).
\end{equation}
Similarly, setting $N=2$ in equation (\ref{Tk}) gives
\begin{subequations}
\begin{align}
&\pi_1T_1 \\
&=\kappa {\mathcal D}^{-1} \bigg ([1+\kappa G_{22}][G'_{11}\pi_1+G'_{12}\pi_2 - G_1']-\kappa G_{12} [G'_{21}\pi_1+G'_{22}\pi_2 - G_2']\bigg),\nonumber\\
&\pi_2T_2 \\
&=\kappa {\mathcal D}^{-1} \bigg([1+\kappa G_{11}][G'_{21}\pi_1+G'_{22}\pi_2 - G_2'] -\kappa G_{21} [G'_{11}\pi_1+G'_{12}\pi_2 - G_1']\bigg ),\nonumber
\end{align}
\end{subequations}
where
\begin{equation}
G'_{kl}=G_1(x_k|x_l),\quad G'_k=G_1(x_k|0).
\end{equation}

\begin{figure}[t!]
\raggedleft
\includegraphics[width=8cm]{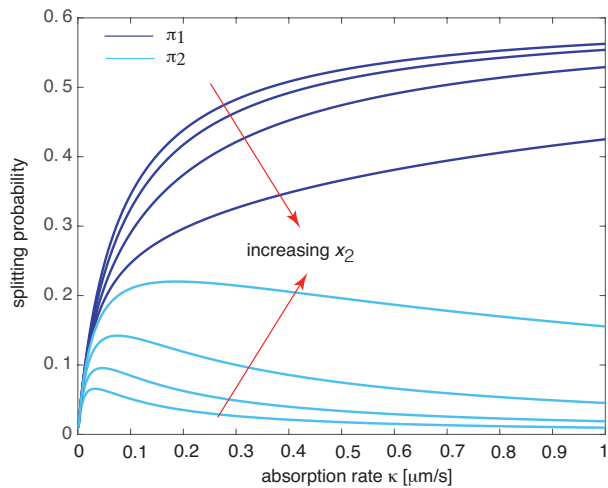} 
\caption{Pair of synaptic targets. Plot of splitting probabilities $\pi_1,\pi_2$ for a pair of synaptic targets at positions $x_1,x_2$. Plots are shown for various positions $x_2=6,10,15,20$ and $x_1=5$. Other parameter values are $D=1$, $v=0.1$ and $L=100$.}
\label{fig6}
\end{figure}

\begin{figure}[b!]
\raggedleft
\includegraphics[width=8cm]{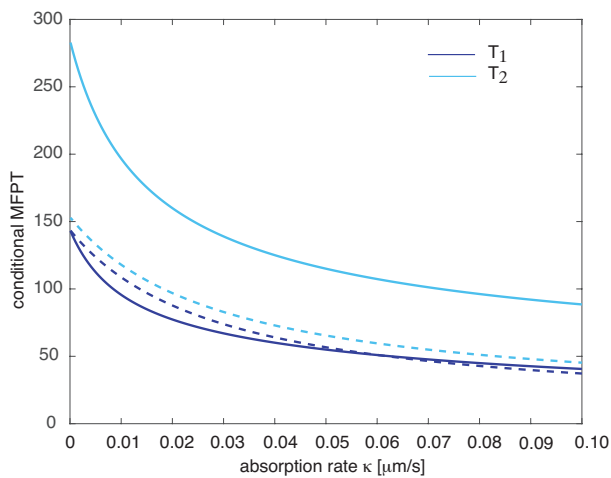} 
\caption{Pair of synaptic targets. Plot of conditional MFPTs $T_1,T_2$ for a pair of synaptic targets at positions $x_1,x_2$. Plots are shown for $x_2=6$ (dashed curves) and $x_2=20$ with $x_1=5$. Other parameter values are $D=1$, $v=0.1$ and $L=100$.}
\label{fig7}
\end{figure}

Example plots of $\pi_k$ and $T_k$, $k=1,2$ are shown in Figs. \ref{fig6} and \ref{fig7}, respectively. A number of observations can be made. First, in the slow absorption limit, the splitting probability and conditional MFPT of the first target become insensitive to the location of the second target, consistent with our previous analysis. Second, $\pi_1$ is a monotonically increasing function of the absorption rate $\kappa$ with $\pi_1\rightarrow 1$ as $\kappa\rightarrow \infty$. On the other hand, $\pi_2$ is a non-monotonic function of $\kappa$ since $\pi_2\rightarrow 0$ as $\kappa \rightarrow \infty$.  This is a consequence of the fact that the first target captures the vesicle with probability one in the fast absorption limit. Note that for finite $\kappa$ we have $\pi_1+\pi_2 <1$ since there is a non-zero probability of absorption at the distal end $x=L$. Finally, the conditional MFPTs are monotonically decreasing functions of $\kappa$.

\setcounter{equation}{0}
 
\section{Mapping to a G/M/$\infty$ queue}

\begin{figure}[b!]
\raggedleft 
\includegraphics[width=10cm]{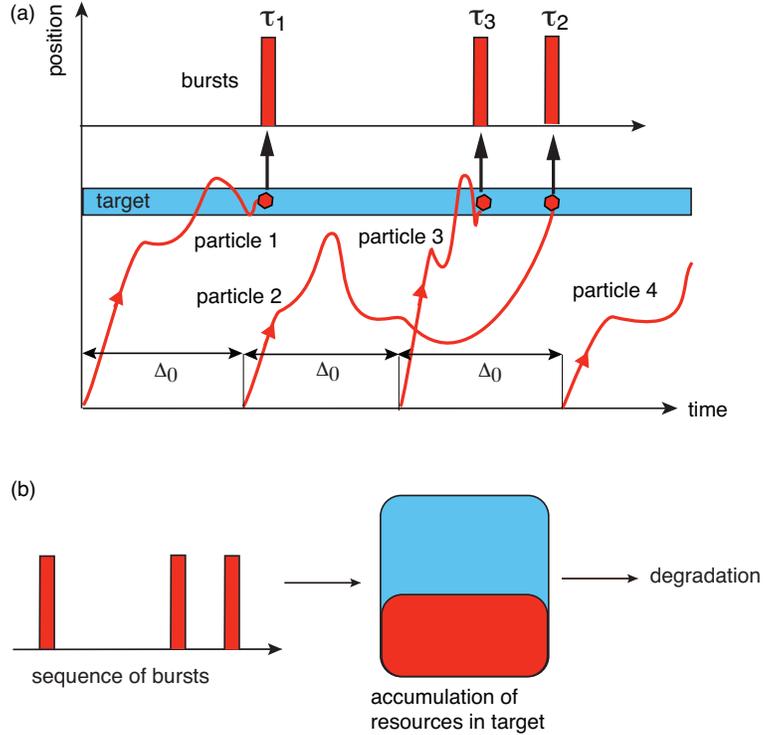} 
\caption{Multiparticle search-and-capture for a single target. (a) Sample particle trajectories. The $j$-th particle starts its search at time $t_j=(j-1)\Delta_0$ and is captured by the target at time $\tau_j=t_j+{\mathcal T}_j(\x_0)$. (b) The random sequence of burst events results in an accumulation of resources within the target, which is counteracted by degradation at some rate $\gamma$.}
\label{fig8}
\end{figure}

In this section we extend the single particle analysis to the case of multiple particles in order to determine the accumulation of synaptic resources along an axon due to the competition between the transport/delivery of cargo and degradation, see Fig. \ref{fig8}. In particular, we show how the statistics of a single search-and-capture model can be incorporated into an infinite server queue, where vesicles represent customer batches, synapses correspond to service stations, and degradation signals the exit of a customer.

\subsection{Multiple search-and-capture events}

Let $t_j$ denote the time of the $j$-th insertion event, $j=1,2,\ldots$, with $t_1=0$. We will assume that the inter-insertion times $\Delta_j=t_{j+1}-t_j$ are generated from a waiting time density $\psi(\Delta)$ with finite mean $\Delta_0$, and that a single particle is inserted each time. The particular case of periodic insertion considered in \cite{Kim19}, $\psi(\Delta)=\delta(\Delta-\Delta_0)$, is illustrated in Fig. \ref{fig8}(a) for a single target. Note that this rule could be further generalized in at least two ways. First, the number of particles injected at time $t_j$ could itself be a random variable $M_j$. Second, the total number of particles in the compartmental pool could be bounded (finite capacity pool). The latter would significantly complicate the analysis, since one would need to keep track of the total number of particles that have been inserted up to time $t$, including any particles that have been recycled to the pool.

Denote the synaptic target that receives the $j$-th vesicle by $k_j$ and define $\tau_j$ to be the time at which the $j$-th particle is captured by the target and delivers its cargo ($j$-th burst event). It follows that
\begin{equation}
\label{tauj}
\tau_j=t_j+{\mathcal T}_{j,k_j}\quad j\geq 1,
\end{equation}
where ${\mathcal T}_{j,k_j}$ is the FPT for the $j$-th particle to find the target $k_j$.
It is important to note that although the insertion times are ordered, $t_{j}<t_{j+1}$ for $j \geq 1$, there is no guarantee that the burst times are also ordered. That is, the condition $\tau_i<\tau_j$ for $i<j$ need not hold. For example, in Fig. \ref{fig8}(a) we see that $\tau_3<\tau_2$. Suppose that a vesicle is delivered to a given target $k$ at the sequence of times $\tau_{j_1,k}$, $\tau_{j_2,k}$ etc. That is, the $n$-th vesicle is delivered to the given target by the particle labeled $j_n$. Consider the difference equation
\begin{equation}
\tau_{j_{n+1,k}}-\tau_{j_{n,k}} =t_{j_{n+1}}-t_{j_n} +{\mathcal T}_{j_{n+1},k} -{\mathcal T}_{j_n,k} .
\end{equation}
Taking expectations of both sides shows that
\begin{align}
\E[\tau_{j_{n+1},k}]-\E[\tau_{j_n,k}]&=\E[t_{j_{n+1}}]-\E[t_{j_n}]+\E[{\mathcal T}_{j_{n+1},k} ]-\E[{\mathcal T}_{j_{n},k} ]\nonumber \\
&=\E[t_{j_{n+1}}]-\E[t_{j_n}].\end{align}
We have used the fact that the search particles are independent and identical so $\E[{\mathcal T}_{j,k} ]=T_k$ independently of $j$. It follows that the mean inter-burst interval $\Delta_k$ to a given target $k$ is independent of the MFPT $T_k$. On the other hand, it does depend on the splitting probability $\pi_k$, since
\begin{align}
\Delta_k\equiv \lim_{N\rightarrow \infty}\frac{1}{N} \sum_{n=1}^N[\E[\tau_{j_{n+1},k}]-\E[\tau_{j_n,k}]]=\frac{\Delta_0}{\pi_k}.
\end{align}
That is, the mean time between particle injections is $\Delta_0$ and only a fraction $\pi_k$ is delivered to the $k$-th target, so that $\Delta_0/\pi_k$ is the expected time separating the injection of particles $j_n$ and $j_{n+1}$.

\subsection{Renewal equation for the Binomial moments}
We now show how to map the accumulation of synaptic resources in our axon transport model to a G/M/$\infty$ queue, generalizing our previous analysis of cytoneme-based morphogenesis \cite{Kim19}. \pcb{We will assume that each vesicle contains $C$ resources (eg. proteins or other macromolecules), each of which degrades (is utilized) independently.} Let $N_k(t)$ be the number of resources within the $k$-th target at time $t$ that have not yet degraded.
In terms of the sequence of capture times $\tau_i$, we can write
\begin{equation} \label{Nt}
	N_k(t) = \sum_{j\geq 1} \chi(t - \tau_j) \delta_{k_j,k},
\end{equation}
where $\delta_{j,k}=1$ if $j=k$ and is zero otherwise, and
\begin{equation}
	\chi(t - \tau_j) = \sum_{d= 1}^{C} I(t- \tau_j, S_{jd})
\end{equation}
for
\begin{equation}
	I(t-\tau_j,S_{jd}) = \left\{\begin{array}{cc}
	0 & \mbox{if } t - \tau_j<0 \\ 
	1 & \mbox{if } 0 \leq t -\tau_j\leq S_{jd} \\
	0 & \mbox{if } t -\tau_j > S_{jd}
	\end{array}\right. .
	\label{Nt3}
\end{equation}
Here $S_{jd}$, $d=1,\ldots,\Delta$, is the service (degradation) time of the $d$th resource element (protein) of the vesicle delivered by the $j$th particle. 

\pcb{The interpretation of equations (\ref{Nt})--(\ref{Nt3}) is as follows. In the absence of degradation, and assuming $N_k(0)=0$, the number of resources within the $k$-th target at time $t$ would simply be the number of packets or vesicles delivered to that target in the interval $[0,t]$ multiplied by the size $C$ of each packet. In terms of the Heaviside function $\Theta$, we would have
\begin{equation}
N_k(t)=C\sum_{j\geq 1} \Theta(t-\tau_j) \delta_{k_j,k}.
\end{equation}
The Heaviside function ensures that we only count capture times that occur within the interval $[0,t]$ and the Kronecker delta only counts the subset of these that involve the $k$-th target. However, when degradation is included, we have to take into account the random loss of resources delivered to the target. Since the $C$ resources delivered at time $\tau_j$ degrade independently of each other, we can assign a time $S_{jd}$ to each of the resources labeled $d=1,\ldots,C$ such that  the $d$-th resource degrades at time $\tau_j+S_{jd}$. It follows that we have to replace the term $C \Theta(t-\tau_j)$ by the sum 
\[\chi(t - \tau_j) =\sum_{d=1}^C \Theta (t-\tau_j) \Theta (\tau_j+S_{jd}-t).\]
This ensures that the given vesicle is captured before time $t$ and we only count resources that haven't yet degraded.
}

Given $N_k(t)$, we define the binomial moments 
\begin{equation}
\label{moments}
	B_{r,k}(t) = \sum_{l=r}^\infty \frac{l!}{(l-r)!r!} \mathbb{P}[N_k(t) = l], \quad r = 1,2,\cdots.
\end{equation}
Introducing the generating function
\begin{equation}
	G_k(z,t) = \sum_{l=0}^\infty z^l \mathbb{P}[N_k(t) = l],\quad t\geq 0,
\end{equation}
we have
\begin{equation}
\label{moments2}
	B_{r,k}(t) = \frac{1}{r!} \left. \frac{d^r G_k(z,t)}{dz^r} \right|_{z = 1}.
\end{equation}
Assuming that the target is empty at time $t = 0$ ($N_k(0)=0$), 
we derive a renewal equation for the generating function $G_k(z,t)$. 
Since the particles are independent, we can decompose equation (\ref{Nt}) as
\begin{equation}
	N_k(t) = \chi(t - \tau_1)\delta_{k_1,k} + \Theta(t-t_2) N_k^*(t- t_2),
\end{equation} 
where $\tau_1$ is the capture time of the first particle injected at $t_1=0$, $t_2$ is the injection time of the second particle, and $N_k^*(t)$ is the accumulation of resources due to all particles but the first. The main step in deriving a renewal equation is to observe that
$N_k^*(t)$ has the same probability distribution as $N_k(t)$. Moreover, $\chi(t - \tau_1)$ and $\Theta(t-t_2) N^*(t-t_2)$ are statistically independent. 
Conditioning on the first arrival time $\tau_1 = \mathcal{T}_1(\x_0) = y$, the target identity $k_1=k$, and the second injection time $t_2=y'$, we have
\begin{align} \label{gzt}
	g(z,t,y,y',k) &\equiv \mathbb{E}[z^{N_k(t)} | \tau_1 = y,t_2=y',k_1=k] \\
	&= \mathbb{E}[z^{\Theta(t-y')N_k^*(t-y')}|t_2=y'] \,\mathbb{E}[z^{\chi(t - y)\delta_{k_1,k}}| \tau_1 = y,k_1=k]. \nonumber
\end{align}
In addition, if $t>y$ then
\begin{equation}
	\mathbb{P}[I(t-y,S_{1d}) =l] = [1- H(t-y)]\delta_{l,1} + H(t-y) \delta_{l,0},
\end{equation}
where $H(t)$ is the service time distribution. Hence,
\begin{equation}
	\sum_{l=0,1} z^l \P[I(t-y,S_{1d}) = l] = z + (1-z)H(t-y)
\end{equation}
for $t>y$.
Since $I(t-y,S_{1d})$ for $d=1,2,\cdots, C$ are independent and identically distributed, the total expectation theorem yields 
\begin{align}
	\overline{G}_k(z,t) &:= \mathbb{E}[z^{\chi(t-\tau_1)\delta_{k_1,k}}] = \mathbb{E}\left[\mathbb{E}[z^{\chi(t-\tau_1)\delta_{k_1,k}}|\tau_1=y,k_1=k]\right]\nonumber \\ &=\mathbb{E}\left[\prod_{d=1}^{C}\mathbb{E}[z^{I(t-y,S_{1d})\delta_{k_1,k}}|\tau_1=y,k_1=k]\right] \nonumber \\
 	&=\int_0^t [z + (1-z) H(t-y)]^C dF_k(y) + \int_t^\infty dF_k(y) +\sum_{k'\neq k}\pi_{k'},
\end{align}
where $dF_k(y)=  J_k(t)dy$. The first two terms on the right-hand side are the contributions from all events such that $k_1=k$ and either $y<t$ or $y>t$; the last term represents the contributions from the remaining events ($k_1\neq k$).
Now applying the total expectation theorem to equation (\ref{gzt}) gives
\begin{align} 
	G_k(z,t) &= \mathbb{E}[z^{N_k(t)}] = \mathbb{E}[g(z,t,y,y',k)] \nonumber \\
	&= \E\left [ \mathbb{E}[z^{\Theta(t-y')N_k^*(t-y')}|t_2=y'] \right ]\cdot \mathbb{E} \left[\mathbb{E}[z^{\chi(t - y)\delta_{k_1,k}}| \tau_1 = y,k_1=k] \right] \nonumber\\
	&=\left ( \int_0^t G_k(z,t-y')  \psi(y')dy' \right )\overline{G}_k(z,t),
\label{Gzt2}
\end{align}
where $\psi$ is the waiting time density for particle insertion.

One can now obtain an iterative equation for the binomial moments by differentiating equation (\ref{Gzt2}) with respect to $z$ and using equation (\ref{moments2}). That is,
\begin{align} 
	B_{r,k}(t) &= \frac{1}{r!} \left. \frac{d^r G_k(z,t)}{dz^r} \right|_{z = 1} \\
	&=\frac{1}{r!}\sum_{l=0}^r {r \choose l} \left ( \int_0^t  \left. \frac{d^{r-l}G_k(z,t-y') }{dz^{r-1}} \right|_{z = 1}\psi(y')dy' \right ) \left. \frac{d^{l}\overline{G}_k(z,t)}{dz^l}\right|_{z = 1}.\nonumber
\label{Gzt3}
\end{align}
Since $H(t)=1-\e^{-\gamma t}$ and
\[
	\left.\frac{d^l}{dz^l} [z + (1-z) H(t-y)]^{C} \right|_{z=1} = \left\{\begin{array}{cc}
	 \frac{\displaystyle {C}!}{\displaystyle (C-l)!} [1 - H(t-y)]^l & \mbox{if } C \geq l \\
	0 & \mbox{if }C< l
	\end{array}\right.,
\]
it follows that
\begin{equation}
 \left. \frac{d^{l}\overline{G}_k(z,t)}{dz^l}\right|_{z = 1}= \frac{\displaystyle {C}!}{\displaystyle (C-l)!}\mathcal{H}_{l,k}(t),
\end{equation}
where $\mathcal{H}_{0,k}(t) = 1$ and
\begin{equation}
\label{Hl}
	\mathcal{H}_{l,k}(t) = \int_0^t e^{-\gamma l (t-y)} dF_k(y), \quad l = 1,2,\cdots.
\end{equation}
On the other hand, from the definition of the Binomial moments,
\begin{equation}
	\frac{1}{r!}  {r \choose l} \int_0^t  \left. \frac{d^{r-l}G_k(z,t-y') }{dz^{r-1}} \right|_{z = 1}\psi(y')dy' =
	\frac{1}{l!}\int_0^t  B_{r-l,k}(t-y')\psi(y')dy' .
\end{equation}
We thus obtain the integral equation
\begin{equation} \label{iterB}
	B_{r,k}(t) = {C \choose r} \mathcal{H}_{r,k}(t) + \sum_{l=0}^{r-1} {C \choose l} \left (\int_0^t B_{r-l,k}(t-y') \psi(y')dy' \right )\mathcal{H}_{l,k}(t) .
\end{equation}

\section{Calculation of the mean and variance}

\pcb{Recall that the main goal of our analysis is to take into account the stochastic nature of synaptic resource accumulation, which cannot be captured using population models. That is, population models only determine the mean distribution of resources along an axon. In the discrete particle model the mean distribution is given by} the first-order moments $B_{1,k}(t)$, which satisfy the renewal equation
\begin{equation} \label{iterB1}
	B_{1,k}(t) = C\int_0^t e^{-\gamma  (t-y)} dF_k(y)+ \int_0^t B_{1,k}(t-y') \psi(y')dy'.
\end{equation}
The integral equation can be solved using Laplace transforms, after making the substitution $dF_k(y) =  J_k(y) dy$. That is,
\begin{equation}
	\widetilde{B}_{1,k}(s) = \widetilde{\psi}(s) \widetilde{B}_{1,k}(s) +C\widetilde{{\mathcal H}}_{1,k}(s),\quad \widetilde{{\mathcal H}}_{1,k}(s)= \frac{ \widetilde{J}_k(s)}{\gamma + s}.
\end{equation}
Solving for $\widetilde{B}_1(s)$ then gives
\begin{equation}
\label{B1s}
	\widetilde{B}_{1,k}(s) = \frac{C}{\gamma + s} \frac{  \widetilde{J}_k(s)}{1 - \widetilde{\psi}(s)}.
\end{equation}
Using the fact that $B_1^*\equiv \lim_{t \to \infty} B_1(t) = \lim_{s \to 0} s \widetilde{B}_1(s)$ and using l'Hospital's rule, we obtain the expression for the steady-state first moment 
\begin{align}
	\langle N_k \rangle & =B_{1,k}^*= \frac{C}{\gamma}\lim_{s \to 0} \frac{s\widetilde{J}_k(s)}{1- \widetilde{\psi}(s)} =- \frac{C}{\gamma}\lim_{s \to 0} \frac{\widetilde{J}_k(s)}{ \widetilde{\psi}'(s)}  = \frac{C\pi_k }{\gamma \Delta_0}, \label{EN}
\end{align}
where $\Delta_0=\int_0^{\infty}\psi(\Delta)\Delta\, d\Delta$ is the mean inter-insertion time and $\pi_k=\widetilde{J}_k(0)$ is the splitting probability. Note that the first moment $\langle N_k \rangle$ in (\ref{EN}) only depends on the mean rate of resource delivery, $C\pi_k /\Delta_0$, divided by the mean degradation rate. It does not depend on the FPT statistics of the search-and-capture process, which is a major difference from sequential search-and-capture \cite{Kim19}. Within the context of queuing theory, equation (\ref{EN}) can be interpreted as a version of Little's law \cite{Little61}, which states that the average number of customers in a stationary system is equal to the long term average effective arrival rate multiplied by the average time that a customer spends in the system. \pcb{One would expect the spatial ($k$-dependent) variation of $\langle N_k\rangle$, as determined by the splitting probabilities $\pi_k$, to be consistent with the steady-state concentration profile obtained using the population model of section 2.1. This is indeed found to be the case. For example, compare the variation of $\Pi(x)$ in Fig. \ref{fig4} with the concentration profile for small $\kappa$ in Fig. \ref{fig3}.}

\pcb{The advantage of the discrete particle model is that it also allows us to determine the size of fluctuations about the mean number of resources. We will illustrate this by calculating the second-order Binomial moments, which yield the variance of the resource distribution. Note, however, that the analysis of higher-order moments is more complicated} due to the presence of terms involving products of time-dependent functions in equation (\ref{iterB}). Setting $r=2$ in equation (\ref{iterB}) gives
\begin{equation} \label{iterB2}
	B_{2,k}(t) ={C \choose 2} \mathcal{H}_{2,k}(t) + \int_0^t B_{2,k}(t-y') \psi(y')dy' 
	+C{\mathcal H}_{1,k}(t) \int_0^t B_{1}(t-y') \psi(y')dy' .
\end{equation}
Squaring both sides of equation (\ref{iterB1}) implies that
\begin{align*}
 2C{\mathcal H}_{1,k}(t) \int_0^t B_{1,k}(t-y') \psi(y')dy' &= B_{1,k}(t)^2-C^2{\mathcal H}_{1,k}(t) ^2\\
 &\quad -\left \{\int_0^t B_{1,k}(t-y') \psi(y')dy' \right \}^2.
\end{align*}
Setting
\[{\mathcal B}_{2,k}(t)=B_{2,k}(t)-\frac{1}{2}B_{1,k}(t)^2\]
and rearranging gives
\begin{align}
 \label{iterB2a}
 {\mathcal B}_{2,k}(t)-\int_0^t{\mathcal B}_{2,k}(t-y){\psi}(y)dy={C \choose 2} \mathcal{H}_{2,k}(t) -\frac{C^2}{2}{\mathcal H}_{1,k}(t) ^2+\frac{1}{2}{\mathcal M}_{1,k}(t),
\end{align}
where
\begin{equation}
\label{M}
 {\mathcal M}_{1,k}(t)= \int_0^t B_{1,k}(t-y')^2 \psi(y')dy' - \left \{\int_0^t B_{1,k}(t-y') \psi(y')dy' \right \}^2.
\end{equation}
Laplace transforming equation (\ref{iterB2a}),
\begin{equation}
	\widetilde{{\mathcal B}}_{2,k}(s) - \widetilde{\psi}(s) \widetilde{{\mathcal B}}_{2,k}(s) ={C \choose 2} \frac{\widetilde{J}_k(s)}{2\gamma + s}-\frac{C^2}{2}\widetilde{\mathcal{H}_{1,k}^2}(s)+\frac{1}{2} \widetilde{\mathcal M}_{1,k}(s),
\end{equation}
and solving for $\widetilde{{\mathcal B}}_{2,k}(s)$ we obtain the result
\begin{equation}
	\widetilde{{\mathcal B}}_{2,k}(s) = {C \choose 2} \frac{1}{2\gamma +s}\frac{\widetilde{J}_k(s)}{1 - \widetilde{\psi}(s)}+\frac{1}{2} \frac{\widetilde{\mathcal M}_{1,k}(s)-C^2\widetilde{\mathcal{H}_{1,k}^2}(s)}{1-\widetilde{\psi}(s)}.
\end{equation}
The steady-state second moment thus takes the form
\begin{align}
B_{2,k}^*&=  \frac{{B_{1,k}^*}^2}{2}+{C \choose 2}\frac{1}{2\gamma}\lim_{s \to 0} \frac{s\widetilde{J}_k(s)}{1- \widetilde{\psi}(s)} +\frac{1}{2} \lim_{s \to 0}\frac{s[\widetilde{\mathcal M}_{1,k}(s)-C^2\widetilde{\mathcal{H}_{1,k}^2}(s)]}{1-\widetilde{\psi}(s)}\nonumber \\
 &= \frac{{B_{1,k}^*}^2}{2}- {C \choose 2}\frac{1}{2\gamma}\lim_{s \to 0} \frac{\widetilde{J}_k(s)}{ \widetilde{\psi}'(s)}-\frac{1}{2} \lim_{s \to 0}\frac{\widetilde{\mathcal M}_{1,k}(s)-C^2\widetilde{\mathcal{H}_{1,k}^2}(s)}{\widetilde{\psi}'(s)} \nonumber \\
 & = \frac{{B_{1,k}^*}^2}{2}+{C \choose 2} \frac{\pi_k}{2\gamma \Delta_0}+\frac{1}{2}\frac{\widetilde{\mathcal M}_{1,k}(0)-C^2\widetilde{\mathcal{H}_{1,k}^2}(0)}{\Delta_0}. \label{EN2}
\end{align}
Using the identity
\begin{equation}
\langle N_k^2\rangle -\langle N_k\rangle^2=2B_{2,k}^*+B_{1,k}^*-{B_{1,k}^*}^2,
\end{equation}
we find that the variance is
\begin{align}
\label{var}
\mbox{Var}[N_k]=\frac{C+1}{2} \langle N_k\rangle +\frac{\widetilde{\mathcal M}_{1,k}(0)-C^2\widetilde{\mathcal{H}_{1,k}^2}(0)}{\Delta_0}.
\end{align}

Further simplification can be obtained in the special case of a periodic insertion rule. In particular, taking $\psi(\Delta)=\delta(\Delta -\Delta_0)$, equation (\ref{M}) becomes
\begin{equation*}
 {\mathcal M}_{1,k}(t)= \int_0^t B_{1,k}(t-y')^2 \delta(y'-\Delta_0)dy' - \left \{\int_0^t B_{1,k}(t-y')\delta(y'-\Delta_0)dy' \right \}^2=0.
\end{equation*}
Hence, for periodic insertion we just have to evaluate the Laplace transform of ${\mathcal H}_{1,k}(t)^2$. The latter takes the form
\begin{align*}
	\widetilde{\mathcal{H}_{1,k}^2}(s) &=\int_0^{\infty}\e^{-st} H_{1,k}(t)^2dt\\
	& = \int_0^{\infty}dt\, \e^{-st}\int_0^t dy \, \e^{-\gamma(t-y}J_k(y)\int_0^t dy' \, \e^{-\gamma(t-y')}J_k(y')\\
	&= \int_0^{\infty}dt\,  \int_0^{\infty}dy\,  \int_0^{\infty}dy'\, \e^{-(2\gamma+s)t} \e^{\gamma(y+y')}J_k(y)J_k(y') \Theta(t-y)\Theta(t-y').
\end{align*}
We can partition the integral into the two cases $y<y'$ and $y>y'$. These two cases yield the same result by symmetry and interchange of $y$ and $y'$. Hence
\begin{align*}
	\widetilde{\mathcal{H}_{1,k}^2}(s)	&=2 \int_0^{\infty}dt\,  \int_0^{\infty}dy\,  \int_0^{\infty}dy'\, \e^{-(2\gamma+s)t} \e^{\gamma(y+y')}J_k(y)J_k(y') \Theta(t-y')\Theta(y'-y)\\
 &=2\int_0^{\infty}dy  \int_{y}^{\infty}dy' \int_{y'}^{\infty}dt \, \e^{-(2\gamma+s)t} \e^{\gamma(y+y')}J_k(y)J_k(y') \\
 &=\frac{2}{2\gamma+s} \int_0^{\infty}dy \int_{y}^{\infty}dy'\, \e^{-\gamma(y'-y)-sy'}J_k(y)J_k(y').
\end{align*}
Setting $s=0$ then gives
\begin{subequations}
\label{app1}
\begin{align}
	\widetilde{\mathcal{H}_{1,k}^2}(0) &=\frac{A_k(\gamma)}{\gamma},\\ A_k(\gamma)&= \int_0^\infty e^{-\gamma y'} \int_0^{\infty}  J_k(y) J_k(y+y') dy dy'.
	\end{align}
\end{subequations}
Finally, setting $\widetilde{\mathcal M}_{1,k}(0)=0$ in equation (\ref{var}) and substituting for $\widetilde{\mathcal{H}_1^2}(0)$ using (\ref{app1}a) gives
\begin{align}
\label{var2}
\mbox{Var}[N_k]=\langle N_k\rangle \left \{ \frac{C+1}{2} -\frac{CA_k(\gamma)}{\pi_k} \right \}.
\end{align}

One of the immediate consequences of equation (\ref{var2}) is that the corresponding Fano factor $\Sigma_k$, which is the ratio of the variance to the mean, is independent of the insertion period $\Delta_0$:
\begin{equation} 
	\Sigma_k := \frac{\langle N_k^2 \rangle - \langle N_k \rangle^2}{\langle N_k \rangle} =\frac{C+1}{2} -\frac{CA_k(\gamma)}{\pi_k} .
	\label{ff}
\end{equation}
Differentiating equation (\ref{app1}b) with respect to $\gamma$ shows that $dA_k(\gamma)/d\gamma <0$ for all $\gamma$, which means that $A_k(\gamma)$ is a monotonically decreasing function of $\gamma$ and $\Sigma_k$ is a monotonically increasing function of $\gamma$. Moreover, in the fast degradation limit, $\gamma \rightarrow \infty$, we see that $A_k(\gamma)\rightarrow 0$ and hence 
\begin{equation}
\Sigma_k\rightarrow \frac{C+1}{2}\quad \mbox{as} \quad \gamma \rightarrow \infty.
\end{equation}
 In order to determine the behavior in the limit $\gamma \rightarrow 0$, we first note that $J_k(t)=\pi_kf_k(t)$, where $f_k(t)$ is the conditional FPT density to be captured by the $k$th target. In particular, from equation (\ref{piT}) we have
\begin{equation}
\int_0^{\infty}f_k(t)dt =1,\quad \int_0^{\infty}tf_k(t)dt = T_k,
\end{equation}
where $T_k$ is the corresponding conditional MFPT. Substituting for $J_k$ in the definition of $A_k(\gamma)$, see equation (\ref{app1}), and performing the change of integration variables $y=T_k\xi$, $y'=T_k \eta$ shows that
\begin{equation}
\widehat{A}_k(\gamma)\equiv \frac{A_k(\gamma)}{\pi_k^2}=T_k^2\int_0^\infty e^{-\gamma T_k \eta } \int_0^{\infty}  f_k(T_k \eta) f_k(T_k(\xi+\eta) )d\xi d\eta.
\label{Ak}
\end{equation}
Introduce the rescaled function 
\begin{equation}
g(\xi)=T_k f(T_k\xi)
\end{equation}
such that
\begin{equation}
\label{gg}
	\int_0^\infty g(\xi) d\xi = 1, \quad \int_0^\infty \xi g(\xi) dx = 1.
\end{equation}
We can then rewrite $\widehat{A}_k(\gamma)$ as
\begin{equation}
\widehat{A}_k(\gamma)=   \int_0^\infty e^{-\gamma T_k \xi} \int_0^{\infty}  g( \eta) g( \xi+\eta) d\eta d\xi .
\end{equation}
It now follows that
\begin{eqnarray}
	\lim_{\gamma \rightarrow 0} \widehat{A}_k(\gamma) &=& \int_0^\infty g(\eta) \int_\eta^\infty g(\xi) d\xi  d\eta = - \frac{1}{2}\int_0^\infty \frac{d}{d\eta} \left[\int_\eta^{\infty} g(\xi)d\xi\right]^2 d\eta \nonumber \\
	&=&  \frac{1}{2}\left [ \int_0^{\infty} g(\xi)d\xi\right]^2 d\eta  = \frac{1}{2}. \label{I10}
\end{eqnarray}
Hence,
\begin{equation}
\Sigma_k\rightarrow \frac{C(1-\pi_k)+1}{2}\quad \mbox{as} \quad \gamma \rightarrow 0.
\end{equation}

\begin{figure}[t!]
\raggedleft
\includegraphics[width=8cm]{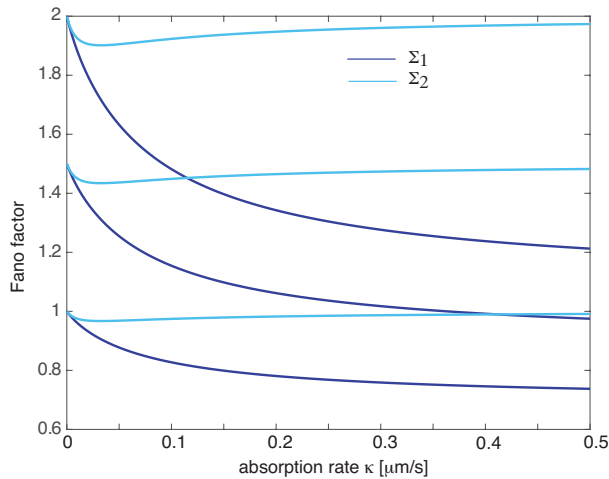} 
\caption{Pair of synaptic targets. Plot of Fano factors $\Sigma_1,\Sigma_2$ in the small $\gamma$ regime for a pair of synaptic targets at positions $x_1=5,x_2=20$. Plots are shown for various vesicle sizes $C$. Other parameter values are $D=1$, $v=0.1$ and $L=100$.}
\label{fig9}
\end{figure}

In summary, we have obtained the following results: (i) the synaptic Fano factors are independent of the insertion rate $\Delta_0^{-1}$; (ii) they are increasing functions of the degradation rate $\gamma$ and the vesicle size $C$; (iii) the Fano factor of the $k$th synaptic target has upper and lower bounds given by
\begin{equation}
\label{boundsT}
\frac{C(1-\pi_k)+1}{2}\leq \Sigma_k \leq \frac{C+1}{2}.
\end{equation}
These bounds imply that if $C=1$ then $\Sigma_k\leq 1$ for all $\gamma$, whereas if $C>1$ then there is a parameter regime in which $\Sigma_k>1$, which means that fluctuations in the number of resources are more bursty than a Poisson process. Now suppose that we combine our results for the Fano factor $\Sigma_k$ with equation (\ref{EN}) for the mean $\langle N_k\rangle$. It can be seen that for fixed $\langle N_k\rangle$, fluctuations can be reduced by simultaneously decreasing $C$ and $\Delta_0$ such that $C/\Delta_0$ is constant. In other words, inserting motor particles with smaller loads more frequently leads to smaller fluctuations. Finally, note that in the small-$\gamma$ regime, more distal synapses have smaller means $\langle N_k\rangle$ and larger Fano factors $\Sigma_k$. The latter is illustrated in Fig. \ref{fig9} for a pair of targets.

\section{Discussion} In this paper we developed a probabilistic framework for investigating the accumulation of resources across an array of synapses in response to the motor-driven axonal transport and delivery of vesicles. There were three major components of the model: (I) The stochastic or periodic insertion of motor particles into the axon; (II) The stochastic dynamics of motor transport along the axon; (III) The uptake of vesicles by synaptic targets. Components II and III for a single particle were formulated as a first passage time problem that determines the statistics of a single search-and-capture event. This was then combined with component I to construct a multiple particle model, which took the form of an infinite server queue. Queuing theory was then used to calculate the steady-state mean and variance of synaptic resource accumulation. 

\pcb{As highlighted throughout the paper, the main reason for considering a discrete particle model of axonal transport rather than the more familiar advection-diffusion model is that the latter cannot account for the discrete and stochastic nature of resource accumulation within an individual synapse. One of the main results of our analysis was to establish that the steady-state Fano factor for the number of resources in a synapse can be significant, particularly when the size $C$ of a vesicle is greater than unity. This means that the time-course of resource accumulation has a strong bursty component, which could interfere with the normal functioning of the synapse, and possibly lead to unreliable synaptic connections between neurons. Since these connections are thought to be the cellular basis of learning and memory, such fluctuations could also be a problem at the organismal level. Indeed, identifying molecular sources of synaptic variability is a topic of general interest within cellular neuroscience \cite{Triller11}. Finally, we note that the mathematical framework developed in this paper provides a basis for exploring a wide range of additional biophysical features, some of which are summarized below.
}

\subsection*{Biophysical models of motor transport} One extension would be to consider a more detailed biophysical model of motor transport (component II). As highlighted in the introduction, the random stop-and-go nature of motor transport can be modeled in terms of a velocity jump process \cite{Newby10a}. For example,
consider a motor-cargo complex that has $N$ distinct velocity states, labeled $n=1,\ldots,N$, with corresponding velocities $v_n$. Take the position $X(t)$ of the complex on a filament track to evolve 
according to the velocity-jump process
\begin{equation}
\label{vjm}
\frac{dX}{dt}=v_{N(t)},
\end{equation}
where the discrete random variable $N(t)\in \{1,\ldots,N\}$ indexes the current velocity state $v_{N(t)}$, and transitions between the velocity states are governed by a discrete Markov process with generator ${\bf 
A}$. Define ${\mathbb P}(x,n,t\mid y,m,0)dx$ as the
joint probability that $x \leq X(t) < x+dx$ and $N(t)=n$ given that
initially the particle was at position $X(0)=y$ and was in state
$N(0)=m$. Setting
\begin{equation}
\label{bbP}
 p_n(x,t) \equiv \sum_{m}\mathbb{P}(x,n,t|0,m,0)\sigma_m,
\end{equation}
with initial condition $p_n(x,0)=\delta (x)\sigma_{n}$, $\sum_{m}\sigma_m=1$, the
 evolution of the probability is described by the differential Chapman-Kolmogorov (CK) equation
\begin{eqnarray}
\label{CKm0}
{\frac{\partial p_n}{\partial t}=-v_n\frac{\partial p_n({x},t)}{\partial x}
+\sum_{n' =1}^N A_{nn'}p_{n'}({x},t) .}
\end{eqnarray}
In the case of bidirectional transport, the velocity states can be partitioned such that $v_n >0$ for $n=1,\ldots, {\mathcal N}$ and $v_n \leq 0$ for $n={\mathcal N}+1,\ldots,N$ with ${\mathcal N} > 0$. 

Suppose that on an appropriate length-scale $L$, the transition rates are fast compared to $v/L$ where $v=\max_n|v_n|$. Performing the rescalings $x\rightarrow x/L$ and $t\rightarrow tv/L$ leads to a non-dimensionalized version of the CK equation
\begin{eqnarray}
\label{CKm}
\frac{\partial p_n}{\partial t}=-v_n\frac{\partial p_n(x,t)}{\partial x}
+\frac{1}{\epsilon}\sum_{n' =1}^N A_{nn'}p_{n'}(x,t),
\end{eqnarray}
with $0< \epsilon \ll 1$. Suppose that the matrix ${\bf A}$ is irreducible with a unique stationary density 
(right eigenvector) $\rho_n$. 
In the limit $\epsilon \rightarrow 0$, $p_n(x,t)\rightarrow \rho_n$ and the motor moves deterministically according to the mean-field equation
\begin{equation}
\label{ratem}
\frac{dx}{dt}=V\equiv \sum_{n=1}^N v_n\rho_n.
\end{equation}
In the regime $0<\epsilon \ll 1$, there are typically a large number of transitions between different motor complex states $n$ while the position $x$ hardly changes at all. This suggests that the system rapidly 
converges to the quasi-steady state $\rho_n $, which will then be perturbed as $x$ slowly evolves. The resulting perturbations can thus be analyzed using a quasi-steady-state diffusion approximation, in which the 
CK equation (\ref{CKm}) is approximated by a Fokker-Planck equation for the total probability density $p(x,t)=\sum_n p_n(x,t)$ \cite{Newby10a}:
\begin{equation}
\label{zFP}
\frac{\partial p}{\partial t}=- V\frac{\partial p}{\partial x} +\epsilon   D\frac{\partial^2 p}{\partial x^2} ,
\end{equation}
with a mean drift $V$ and a diffusion coefficient $D$ given by
\begin{align}
\label{eq7:calD}
D =\sum_{n=1}^NZ_n v_n,
\end{align}
where $Z_n $, $\sum_mZ_m=0$, is the unique solution to
\begin{equation}
\label{eq7:Z}
\sum_{m=1}^N A_{nm} Z_m =[V -v_n]\rho_n .
\end{equation}
Hence, we recover the FP equation used in the single-particle model of section 2, except that now the drift and diffusion terms preserve certain details regarding the underlying biophysics of motor transport due to the dependence of $V$ and $D$ on underlying biophysical parameters. 

\subsection*{Local signaling}  Using a more detailed biophysical transport model means that we could incorporate local inhomogeneities due to chemical signaling, for example. One of the major signaling mechanisms 
 involves microtubule associated proteins (MAPs). These molecules bind to microtubules and effectively modify the free energy landscape of motor-microtubule interactions \cite{Telley09}. For example, tau is a MAP found in the axon of neurons and is known to be a key player in Alzheimer's disease \cite{Kosik86}. Experiments have shown that the 
presence of tau on a microtubule can significantly alter the dynamics of kinesin; specifically, by reducing the rate at which kinesin binds to the microtubule \cite{Vershinin07}. Within the context of velocity jump processes, local variations in tau concentration would lead to $x-$dependent switching rates between the different velocity states. That is, the matrix generator ${\bf A}$ and, hence the drift velocity and diffusivity, become $x$-dependent  \cite{Newby10b,Newby10c}. It is also known that abnormal hyperphosphorylation of tau can disrupt the role of tau in promoting the assembly and stabilization of microtubules, which is thought to be an important step in the progression of Alzheimer disease \cite{Wang15}. It would be interesting in future work to use the queuing modeling framework to investigate the effects of tau signaling on the accumulation of synaptic resources.

\subsection*{Transfer of vesicles to synaptic targets} In this paper we treated each synaptic target as a partially absorbing, point-like sink (component III). Representing each target in terms of a Dirac delta function was possible due to the fact that the axon was modeled as a one-dimensional cable, which meant that the associated one-dimensional Green's function was non-singular. However, this quasi-1D approximation is not appropriate for synapses distributed over a more local region of an axon or dendrite nor for synapses located in the somatic membrane. In such cases one can no longer treat the synapses as point-like, since the corresponding two-dimensional  Green's function has a logarithmic singularity. However, if the synapses are relatively small compared to the search domain then one can use asymptotic methods to solve the first passage time problem for a single particle by extending previous studies \cite{Straube07,Bressloff08,Coombs09,Lindsay16,Bressloff21a} to the case of a partially absorbing target.
Finally, note that the detailed mechanism underlying the transfer of vesicular cargo from motor complexes to synapses is not well understood, although it is likely to involve myosin motors and the local actin cortex. Incorporating such details would require replacing simple partial absorption by a more complicated kinetic scheme \cite{Schumm21}. Such a scheme could also include a more detailed model of how resources are subsequently utilized, beyond simple degradation.

\end{document}